\documentclass[10pt,a4paper,twoside]{article}
\usepackage{baltlat6}
\usepackage{array}
\usepackage{here}
\usepackage{epsfig}
\usepackage{epstopdf} 
\begin{document}
\ \
\vspace{0.5mm}
\setcounter{page}{1}

\titlehead{Baltic Astronomy, vol.\,??, ???--???, 2016}

\titleb{RADIAL VELOCITY DISPERSIONS OF STAR GROUPS IN M~67 \\}

\begin{authorl}
\authorb{S.V. Vereshchagin}{1} and
\authorb{N.V. Chupina}{1}
\end{authorl}

\begin{addressl}
\addressb{1}{Institute of Astronomy of the Russian Academy of Sciences,\\
48 Pyatnitskaya st., Moscow, Russia;
svvs@ya.ru}
\end{addressl}

\submitb{Received: 2016 July; accepted: 2016 July}

\begin{summary} 
The high-precision measurements of the radial velocities ($V_r$), 
which were obtained  for the member stars of M~67, 
are used to calculate the Vr dispersions in earlier found 
stellar groups in the cluster corona. 
The previously detected feature for stars in one of the groups (gr.~60) 
consisting of the almost full identity of spatial velocities was confirmed. 
The possibility of more accurate groups parameters studies in future 
with the help of the GAIA catalogues is discussed.
\end{summary}


\resthead{Radial velocity dispersion of star groups in M~67}
{S.V. Vereshchagin, N.V. Chupina}

\sectionb{1}{INTRODUCTION}

Star group due to the small number of members 
do not show any significant concentration of stars in the sky field plane. 
To find them special algorithm is developed, 
which is based on analyzing the stellar density distributions 
to recognize groups of stars. 
Also, one can search areas with higher concentration of stars 
using observations in X-ray, $H_\alpha$ and the other photometric bands, 
which will be very effective. 
It allows to look inside and behind the molecular gas clouds 
which hide the stars. 
Thus, the observed star distribution is deceptive and becomes more evident 
with the help of suitable filters and special methods. 
Our contribution to this direction of research is 
to detect groups in the corona of open star cluster (OC) M~67 
(Chupina \& Vereshchagin 1998) 
and group 189 near NGC~1977 cluster (Chupina \& Vereshchagin 2000). 
It should be noted that one of the first papers on the topic 
was published almost forty years ago by Latyshev (1977). 
The problems connected with the groups detection is discussed by Mamajek et al. (2015). 

To confirm the reality of existence of the  stellar groups, 
a number of additional checks should be conducted, 
including verification on the random fluctuations of stellar density. 
Also, the differences between parameters of the stars, 
for example, proper motions and radial velocities, 
should be estimated with statistical significance. 
More accurate observation increases the reliability of such tests.

\sectionb{2}{THE SPECTRUM OF VELOCITY FLUCTUATIONS}

M67 is the nearest to us from the old OC (7.0$\pm$0.5 Gyr, Jiaxin e.a. 2015). 
The cluster is high in the disc ($Z=480$~pc), 
moving on an orbit with a large inclination to the disk plane, Wu e.a. (2009). 
These indicate that the dynamic interactions with 
molecular gas clouds of the disk is not occur 
and there is no dissipation of stars from cluster owing to external interactions.

Chupina \& Vereshchagin (1998), used data from Frolov \& Ananyevskaya (1986) 
have detected five stellar groups in the corona of M~67. 
The groups are located in the low density area. 
They include from 5 to 11 stars and have average sizes from 0.5 to 1.2 pc. 
By analysis of proper motions, it has been proven that the groups 
are not random fluctuations of density.

Geller et al. (2015) measured the radial velocities for 13776 stars, 
of which 1278 are candidate members of M~67. 
An average error of $V_r$ is less than 1 km/s, 
instead of the prevalent value of 3 km/s. 
The catalogue has the area size of $2.5^\circ\times1.5^\circ$, 
which is comparable with Frolov \& Ananyevskaya (1986). 
Using high-precision data, 
we rechecked the existence of the groups in M67 by radial velocities. 
Note that Vereshchagin et al. (2014) considered only the central part of the cluster 
and thus obtained different results.
Figure~1 shows the distribution of stars on the celestial sphere 
from the catalogue given by Frolov \& Ananyevskaya (1986) 
and stars of the groups, for which are measured $V_r$ in Geller et al. (2015).


\begin{figure}
\vbox{
\centerline{\psfig{figure=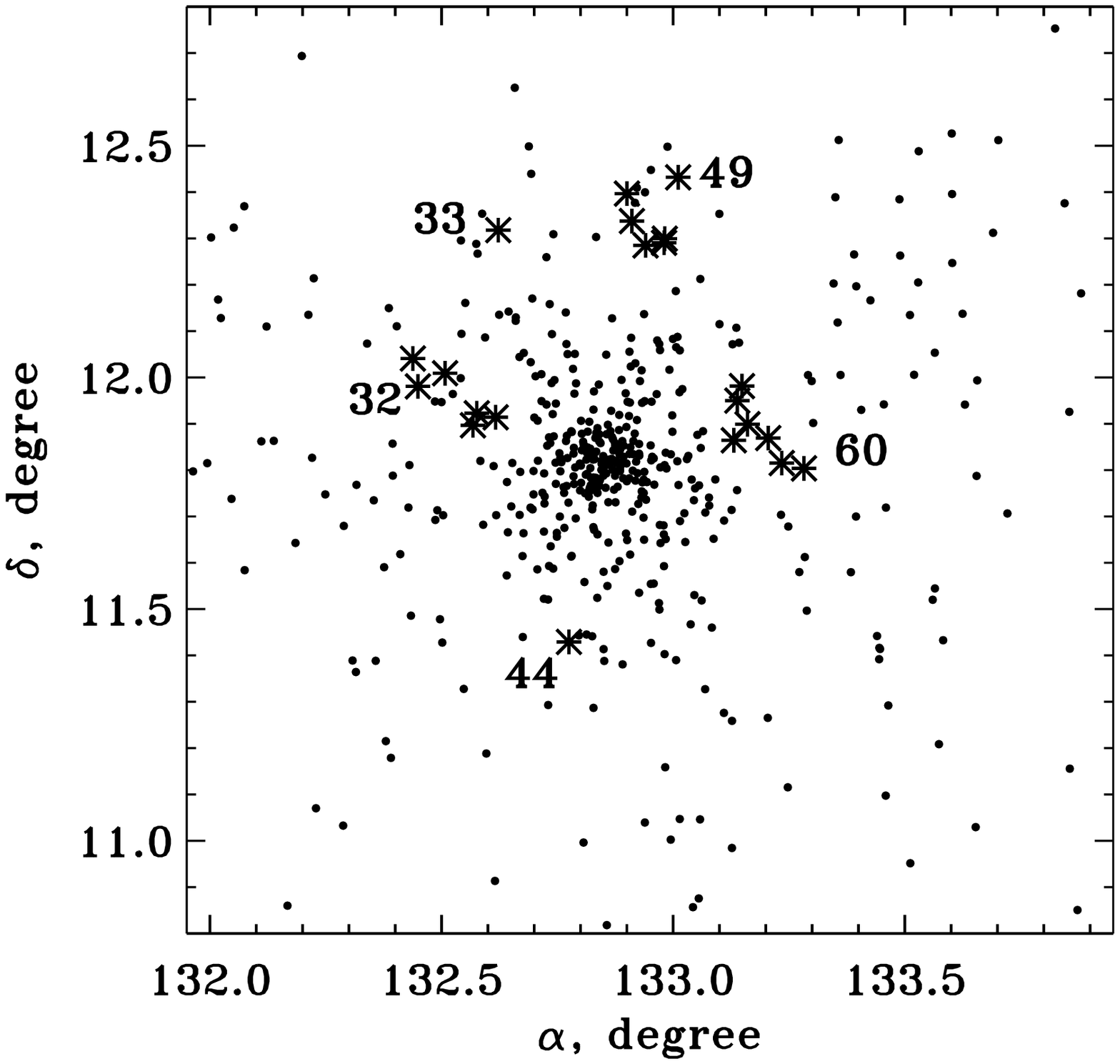,width=80mm,angle=0,clip=}}
\vspace{1mm}
\captionb{1}
{The distribution of the M~67 region stars on the celestial sphere. 
The dots show the stars from Frolov \& Ananyevskaya (1986). 
Stars indicate the stars of the groups 
with radial velocities found in the catalogue by Geller et al. (2015). 
The group numbers are also shown.
}
}
\end{figure}

\sectionb{3}{THE DISPERSION OF RADIAL VELOCITIES}

Geller et al. (2015) list the M~67 membership probabilities 
given by various authors. 
First, we selected stars, for which at least one author have 
non-zero membership probability ($P$), 
then we have taken stars with $P\ge60\%$ by Geller e.a. (2015). 
Our sample consisted of 627 stars. 
According to the sample, the average value of the radial velocity is 
$\langle V_r\rangle =33.58\pm 0.86$~km/s. 
The same we have defined for the groups shown in Figure~1. 
The results are in Table~1. 
Using Fig.1 from Chupina \& Vereshchagin (1998) 
we calculated the average sizes $\langle l\rangle$ of groups, 
which are also in Table~1. 
The first column of Table~1 shows the group number according to Chupina \& Vereshchagin (1998) 
or star number according to Geller et al. (2015). 
The following columns exhibit the $V$-magnitude ($V$), radial velocity ($V_r$) 
and its error ($\varepsilon_{V_r}$) from Geller et al. (2015), 
the calculated average value $\langle V_r\rangle$ and 
mean quadratic deviation $\sigma_{V_r}$ for the groups. 
Columns (7) and (8) have the mean sizes of the groups in $mm$ and in pc. 
Their values are in the range from 1.85 to 3.01~pc. 
Since the groups outlines are irregular in shape, 
we took the average value between its largest and smallest diameters. 
In the group~32, star with  number 4046 by Geller et al. (2015) is discarded. 
For groups 33 and 44, we found only one star 
with measured $V_r$ and the $\sigma_{V_r}$ is not defined.


\begin{table}[!t]
\begin{center}
\vbox{\footnotesize\tabcolsep=3pt
\parbox[c]{124mm}{\baselineskip=10pt
{\smallbf\ \ Table 1.}{\small\
The star groups parameters.
\lstrut}}
\begin{tabular}{cccccccc}
\hline
\noalign{\smallskip}
The group/&V,&$V_r$,&$\varepsilon_{V_r}$,&$\langle V_r\rangle$,&$\sigma_{V_r}$,&$\langle l\rangle$,&$\langle l\rangle$\\
star number&mag&km/s&km/s&km/s&km/s&$mm$&pc\\
\hline
\noalign{\smallskip}
gr. 32&&&&35.9&5.1&12.47&3.01\\
10055&13.52&32.60&0.20&&&&\\
7051&13.34&35.00&0.21&&&&\\
4046&13.35&6.78&0.31&&&&\\
4035&12.86&33.66&0.14&&&&\\
13035&9.98&45.00&0.10&&&&\\
11030&14.18&33.35&0.30&&&&\\
\hline
\noalign{\smallskip}
gr. 33&&&&&&7.66&1.85\\
5066&12.54&35.00&0.06&&&&\\
\hline
\noalign{\smallskip}
gr. 44&&&&&&6.79&1.64\\
5048&13.55&33.62&0.46&&&&\\
\hline
\noalign{\smallskip}
gr. 49&&&&41.0&15.0&12.47&3.01\\
6070&14.36&34.88&0.24&&&&\\
4063&12.96&33.60&0.13&&&&\\
1058&11.34&49.47&0.13&&&&\\
3059&10.87&27.44&0.13&&&&\\
3060&13.01&33.14&0.15&&&&\\
1077&10.02&67.65&0.15&&&&\\
\hline
\noalign{\smallskip}
gr. 60&&&&33.0&0.5&12.47&3.01\\
4034&13.71&33.38&0.16&&&&\\
12038&14.44&32.54&0.30&&&&\\
5041&13.41&33.10&0.24&&&&\\
5039&13.64&33.60&0.14&&&&\\
5043&13.50&32.88&0.19&&&&\\
2046&12.59&32.26&0.26&&&&\\
8052&13.67&33.39&0.29&&&&\\
\hline
\end{tabular}
}
\end{center}
\end{table}

\sectionb{4}{ON STAR GROUPS PHYSICS}

To estimate the velocity dispersion we apply the formula:
$$\sigma_V = \sqrt{\frac{\gamma\cdot M_{cluster}}{R_{cluster}}},$$
where $\gamma$ -- gravitational constant.
The parameters for the statistically stationary cluster with
parameters similar to M~67 are $M_{cluster}=1000M_\odot$ (the 1000 star of the same mass equal to 1 Solar mass), 
$R_{cluster}=100$~pc. 
For M~67, $\sigma_{V_r}=0.14$~km/s. 
For groups containing about 10 stars in a region with radius of 1.5~pc, 
get almost the same value $\sigma_{V_r}=0.12$~km/s. 
Thus, for the stability of a cluster as a whole and 
for internal star groups in cluster corona too, 
it is required to have the dispersion value approximately equal to 0.12 -- 0.14 km/s. 
The observed dispersion given in Table~1 is significantly 
superior to this theoretical value interval, 
even for the group~60, 
which has the minimum value of 0.5~km/s. 
Perhaps this is due to the groups not being in the steady state. 

There are several approaches, which explain this problem. 
In the star systems with a small velocity dispersion 
the mutual gravity can generate paired and multiple correlation 
of the position and velocities of stars. 
Modern calculations show different instabilities 
inside the OC and the appearance of substructures. 
Thus, according to the N-body calculations 
for model of spherical star systems (Polyachenko 2015) 
the cluster may become non-uniform with regions of high density, 
in late stage of its evolution, 
when the age of the cluster exceeds the relaxation time.  

The calculations of the open cluster dynamics 
show the formation of the star groups or streams 
with the similar space trajectories and dynamic characteristics 
(Danilov 2002, 2005; and Danilov \& Leskov 2005). 
Because the formation of stellar groups 
with the similar orbit oscillation 
on the periphery of the cluster (Danilov 2005, 2006) 
can be the force field fluctuations. 
To understand the influence of instability on the dynamics and kinematics 
of the star groups inside the OC, one can analyse 
the phase density fluctuations and their spectra. 
This method gives an indication of the formation 
of the density waves and groups of stars with different 
kinematic and dynamic characteristics (Danilov \& Putkov 2013, 2014; Danilov 2015).

\sectionb{5}{DISCUSSION AND CONCLUSIONS}

5.1. Gr.~60 (see Table~1) stands apart from others, 
it has very small dispersion equal to 0.5 km/s. 
This value is close to the physical limit 
of radial velocities precision for the individual stars. 
This limit is associated with the gas motions 
in the stellar atmospheres (granulation, etc.). 
A similar result for the tangential velocities of this group 
was obtained by Chupina \& Vereshchagin (1998). 
Also this group has a small number of giants and supergiants 
in comparison with the cluster as a whole. 

It can provide information about the time of its formation 
by analyze of the CMD-diagram. 
This result suggests that the stars of gr.60 are physical related. 
Is gr.60 a group of dynamic nature or hierarchical system? 
On the similarities and differences between our groups and hierarchical systems 
we give the following argument. 
The average distance between Mizar and Alcor is about 0.64 pc, 
this is similar to the distance in the groups. 
The Mizar-Alcor hierarchical system has two main components -- 
stars located at a large distance, 
and the other stars in their vicinity. 
In the groups all stars are in average equidistantly. 
But, question of all hierarchical systems are similar to Mizar-Alcor, 
or they may consist of the equidistant stars is still unanswered. 
It can be clarify by the dynamic calculation and 
analysis of the structure of the known systems. 
It is useful to take into account the peculiarities 
of $V_r$ dispersion in hierarchical systems.

5.2. The estimation of significance level. 
We accept the null hypothesis: 
dispersion for the gr.~60 (0.5 km/s) 
and for the cluster as a whole (0.86 km/s) are equal. 
On the criterion of a Fischer-Snedecor 
$F=0.86/0.5=1.72$. 
Degrees of freedom: $k_1=627-1=626$ (whole cluster), 
$k_2=7-1=6$ (gr.~60). 
We choose a significance level equal $\alpha=0.05$. 
The critical value is $F_{cr}(626;6;0.05)=2.1$. 
Thus, $F<F_{cr}$ and the null hypothesis is not rejected. 
This means that the radial velocity dispersion of the group 
and one of the whole cluster are not formally differ. 
Group~60 statistically insignificantly different from the cluster as a whole. 
However, looking at Table~1, 
we see that the $V_r$ dispersions of the other groups (32 and 49) 
are significantly different from both 60 and from the entire cluster. 
Although it should be noted the effect of 
individual stars with $V_r$-value very remote from others. 
The reason for the latter is worth considering in the future. 

5.3. The question about the difference between the groups 
and sparsely populated clusters exist. 
What are the minima OC dimensions? 
Table~2 shows the clusters with the smallest number of their member stars, 
which we have got from MWSC~II given by Kharchenko et al. (2013). 
In Table~2, we include clusters located not far away (within 1 kpc) from the Sun. 
The Table~2 contains the equatorial coordinates on J2000, 
MWSC internal number and the cluster name, 
$Nr_0$, $Nr_1$, $Nr_2$ the numbers of stars inside the angular radius 
of the core ($r_0$), of the central part ($r_1$) and of the cluster ($r_2$). 
The distances from the Sun (dist) and log ages ($\log(t)$) are also included. 
This distance is comparable with M~67. 
Their angular sizes ($r_2$, degree) were converted to linear radii ($r_2$, pc) 
using distances from the Sun (dist). 
From Table~2, we see that size of these clusters and 
the group sizes ($l$ in Table~1) are in good agreement.


\begin{table}[!t]
\begin{center}
\vbox{\footnotesize\tabcolsep=3pt
\parbox[c]{124mm}{\baselineskip=10pt
{\smallbf\ \ Table 2.}{\small\
The sparsely populated clusters from MWSC~II.
\lstrut}}
\begin{tabular}{ccccccccccccc}
\hline
\noalign{\smallskip}
$\alpha$&$\delta$&MWSC&name&$Nr_0$&$Nr_1$&$Nr_2$&$r_0$&$r_1$&$r_2$&$r_2$,&dist,&$\log(t)$\\
\multicolumn{2}{c}{degree}&num&&&&&\multicolumn{3}{c}{degree}&pc&pc&\\
\hline
\noalign{\smallskip}
084.4755&-06.9560&0607&KMS\_35&5&6&7&0.018&0.060&0.095&1.4&636&6.600\\
084.6825&-02.6000&0612&Sigma\_Ori&2&6&9&0.030&0.280&0.400&5.8&390&6.100\\
086.6775&+00.1000&0664&NGC\_2068&0&5&9&0.020&0.075&0.140&2.0&747&6.450\\
093.5100&-09.3750&0807&FSR\_1120&3&7&9&0.020&0.075&0.140&2.0&830&7.300\\
110.9835&-25.3590&1205&FSR\_1293&2&8&10&0.010&0.037&0.080&1.2&962&9.005\\
269.9325&-28.2070&2769&Trumpler\_31&2&6&12&0.018&0.090&0.165&2.4&830&8.635\\
308.4585&+40.1150&3367&Kronberger\_59&3&8&12&0.010&0.045&0.075&1.1&844&8.000\\
110.6550&-05.1150&1194&FSR\_1157&2&8&14&0.010&0.045&0.090&1.3&926&7.700\\
260.1705&-35.8830&2575&vdBergh-&3&9&15&0.015&0.045&0.070&1.0&895&8.450\\
&&&Hagen\_223&&&&&&&&&\\
086.7915&+00.3000&0665&NGC\_2071&2&8&16&0.020&0.090&0.140&2.0&953&6.550\\
268.1250&-30.1050&2727&Basel\_5&1&6&16&0.020&0.120&0.200&2.9&995&8.850\\
264.4725&-08.0850&2650&Mamajek\_2&3&8&17&0.060&0.190&0.310&4.5&174&8.100\\
139.2000&-47.9450&1645&DBSB\_36&1&10&18&0.015&0.070&0.130&1.9&748&7.500\\
328.3725&+47.2650&3571&IC\_5146&1&5&18&0.020&0.080&0.165&2.4&770&6.000\\
130.4700&-79.0450&1535&Mamajek\_1&1&5&19&0.030&0.190&0.420&6.1&112&6.990\\
\hline
\end{tabular}
}
\end{center}
\end{table}

5.4. The proper motion accuracy in the GAIA project 
will be up to 20 microsecond per year ($\varepsilon_\mu =20\;\mu$as/yr) (de Bruijne et al. 2010). 
We converted it into the km/s 
for the distance from the Sun equal to 890 pc (as M~67) 
and got the tangential velocity $V_t=\frac{4.74\mu}{\pi}=0.08$~km/s. 
This is a very small value. 
If, for comparison, take the error equals 1 mas/yr 
as Hipparcos catalog (ESA, 1997) proper motion, 
we obtain the greater value $V_t=4.2$~km/s. 
How many M~67 members will be observed by GAIA? 
We compare stellar surface density of M~67 and GAIA. 
In GAIA, the number of stars vary for the data with known $V_r$, $\mu$, $\pi$. 
GAIA has 24240 star per square degree for stars with known $\mu$ 
and about 5000 stars per square degree for stars with known $\pi$. 
The density of stars in M~67 is much smaller and 
is approximately 318 stars per square degree.
Thus, most likely in GAIA we will find not only the proper motions, 
but also the parallaxes for all stars in the M~67.


\begin{figure}
\vbox{
\centerline{\psfig{figure=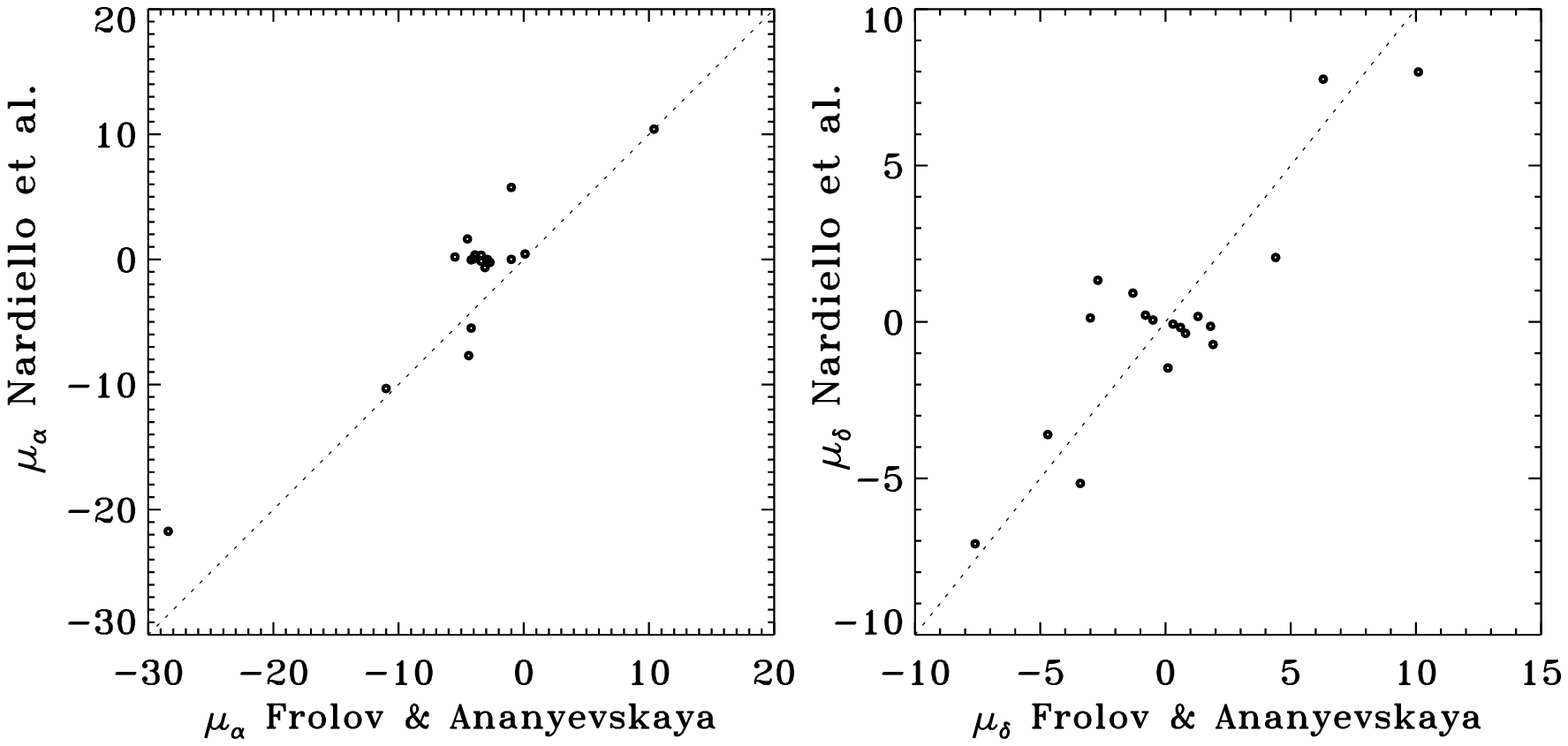,width=120mm,angle=0,clip=}}
\vspace{1mm}
\captionb{2}
{The PM (in mas/yr) comparison for the common groups members stars 
of Frolov \& Ananyevskaya (1986) and Nardiello et al. (2015) catalogues.
}
}
\end{figure}

5.5. In the catalog Frolov \& Ananyevskaya  (1986), 
we found a high-velocity star in the corona. 
Its data: catalog number 1007, $\alpha=133.70202^\circ$, $\delta=+12.51210^\circ$ (J2000), 
$V=11.18$~mag, $\mu_x=-0.3085$, $\mu_y=-0.2211$, arcsec/yr.  
Although the probability of membership is only 3\%, 
it is necessary to understand its nature carefully, 
because there are high-speed stars in M~67. 
In the catalogues by Nardiello et al.(2015) and Geller et al. (2015), 
this star is not available. 
We will continue to study it, 
considering the many interesting features of M~67. 
In particular, in the M~67 some stars with the exoplanets are detected, Brucalassi et al. (2014). 

5.6. The refined proper motions by Nardiello et al. (2015) 
have not yet been used by us. 
They obtained proper motions only for the central part of the M~67 ($34'\times33'$) down to $V\approx 22$ mag.
The region studied by us covered $2^\circ\times 2^\circ$. 
Small area led to a small number of matched stars in our sample. 
With the catalog by Frolov \& Ananyevskaya (1986) identifies 218 stars, 
of which only 14 stars belongs to the groups. 
We matched the Frolov \& Ananyevskaya  (1986) 
and Nardiello et al. (2015) catalogues in Figure~2. 
The points scatter in Figure~2 show proper motion disagreement 
(approximately equal to $3\sigma$) amounted to 10 mas. 
These comparison showed the need for revision of 
our old results with the help of Nardiello et al. (2015) catalogue.

ACKNOWLEDGMENTS. The authors are thankful to Danilov~V.M., D.~Nardiello, 
A.V.~Loctin and Devesh P.~Sariya for their help. 
This work is partly supported by the Russian Foundation for Basic Research 
(RFBR, grant number is 16-52-12027). 
The use of the Simbad database is acknowledged.

\References

\refb Brucalassi A., Pasquini L., Saglia R., Ruiz M.T., Bonifacio P., Bedin L.R., 
Biazzo K., Melo C., Lovis C., Randich S. 2014, A\&A, 561, 9

\refb de Bruijne J., Kohley R., Prusti T. 2010, Space Telescopes and Instrumentation 2010: Optical, Infrared, and Millimeter Wave. Eds. Oschmann J.M., Clampin M.C., MacEwen H.A. Proceedings of the SPIE, V.7731, article id. 77311C, p.15

\refb Chupina N.V., Vereshchagin S.V. 1998, A\&A, 334, 552

\refb Chupina N.V., Vereshchagin S.V. 2000, In Star formation from the small to the large scale. ESLAB symposium; Edited by F. Favata, A. Kaas, and A. Wilson; ESA SP 445; Noordwijk, The Netherlands: ESA, p.347

\refb Danilov V.M. 2002, AZh, 79, 986

\refb Danilov V.M. 2005, AZh, 82, 678

\refb Danilov V.M., Leskov E.V. 2005, AZh, 82, 219

\refb Danilov V.M. 2006, AZh, 83, 393

\refb Danilov V.M., Putkov S.I. 2015,  Astrophys. Bulletin, 70, 71

\refb Danilov V.M., Putkov S.I. 2013, Astrophys. Bulletin, 68, 154

\refb Danilov V.M., Putkov S.I. 2014,  Astrophys. Bulletin, 69, 27

\refb ESA 1997, The Hipparcos and Tycho Catalogues, ESA SP-1200 

\refb Frolov V.N., Ananyevskaya J.K. 1986, Photometry and proper motions in the open cluster M67 (NGC 2682), Cent. Don. Stell., Strasbourg, Catalogue number 5052

\refb Global Astrometric Interferometer for Astrophysics (GAIA), 2016\\ http://www.cosmos.esa.int/web/gaia/

\refb Geller A. M., Latham D.W., and Mathieu R.D. 2015, AJ, 150, 97

\refb Kharchenko N.V., Piskunov A.E., Schilnach E., Roeser S., Scholz R. 2013, A\&A, 558, 53

\refb Latyshev I.N. 1977, Astron. Circ., No. 969,  7

\refb Mamajek E.E. 2015, In Young Stars and Planets, Proc. IAU Symposium No. 314, J.H. Kastner, B. Stelzer and S.A. Metchev, eds.

\refb Nardiello D., Libralato M., Bedin L.R., Piotto G., Ochner P., Cunial A., Borsato L., Granata V. 2016, MNRAS, 455, 3, 2337

\refb Polyachenko E.V. 2015, private message

\refb Vereshchagin S.V., Chupina N.V., Sariya Devesh P., Yadav R.K.S., Brijesh Kumar 2014, New Astronomy, 31, 43

\refb Wang J.-X., Ma J., Wu Z.-Y., Wang S., and Zhou X. 2015, AJ, 150, 61

\refb Wu Z., Zhou X., Ma J. and Du C. 2009, MNRAS, 399, 2146

\end{document}